\documentclass{chi2011}

\usepackage{ucs}
\usepackage[utf8x]{inputenc}

\usepackage{times}
\usepackage{url}
\usepackage{graphics}
\usepackage{color}
\usepackage[pdftex]{hyperref}
\hypersetup{%
pdftitle={Computers Can't Give Credit: How Automatic Attribution Falls Short in an Online Remixing Community},
pdfauthor={Andres Monroy-Hernandez},
pdfkeywords={remixing, online communities, education, computers and children, creativity support tools, end user programming, social computing and social navigation, computer supported cooperative work, computer mediated communication},
bookmarksnumbered,
pdfstartview={FitH},
colorlinks,
citecolor=black,
filecolor=black,
linkcolor=black,
urlcolor=black,
breaklinks=true,
}
\newcommand{\comment}[1]{}
\definecolor{Orange}{rgb}{1,0.5,0}

\begin{document}
\clubpenalty=10000
\widowpenalty = 10000

\setlength{\paperheight}{11in}
\setlength{\paperwidth}{8.5in}
\setlength{\pdfpageheight}{\paperheight}
\setlength{\pdfpagewidth}{\paperwidth}

\toappear{Best Paper Honorable Mention at CHI 2011.}

\title{Computers Can't Give Credit: How Automatic Attribution Falls Short in an Online Remixing Community} 
\numberofauthors{1}
\author{
  \alignauthor Andrés Monroy-Hernández$^{1,2}$, Benjamin Mako Hill$^1$\\ Jazmin Gonzalez-Rivero$^2$, danah boyd$^2$\\
  \affaddr{$^1$Massachusetts Institute of Technology, Cambridge, MA 02139}\\
  \affaddr{$^2$Microsoft Research, Cambridge, MA 02142}\\
  \email{\{amonroy,mako\}@mit.edu} \email{\{t-jazgo,dmb\}@microsoft.com}
}

\maketitle

\sloppy

\begin{abstract}
  In this paper, we explore the role that attribution plays in shaping
  user reactions to content reuse, or remixing, in a large
  user-generated content community. We present two studies using data
  from the Scratch online community -- a social media platform where
  hundreds of thousands of young people share and remix animations and
  video games.  First, we present a quantitative analysis that
  examines the effects of a technological design intervention
  introducing automated attribution of remixes on users' reactions to
  being remixed. We compare this analysis to a parallel examination of
  ``manual'' credit-giving. Second, we present a qualitative analysis
  of twelve in-depth, semi-structured, interviews with Scratch
  participants on the subject of remixing and attribution. Results
  from both studies suggest that automatic \emph{attribution} done by
  technological systems (i.e., the listing of names of contributors)
  plays a role that is distinct from, and less valuable than,
  \emph{credit} which may superficially involve identical information
  but takes on new meaning when it is given by a human remixer. We
  discuss the implications of these findings for the designers of
  online communities and social media platforms.
\end{abstract}





\section{Introduction}

Networked information technologies have changed the way people use and
reuse creative -- and frequently copyrighted -- materials. This change
has generated excitement, and heated debate, among
content-creators, technologists, legal academics, and media scholars.
Media theorist Lev Manovich argues that remixing is an ancient
cultural tradition (e.g., he has suggested that ancient Rome was a
``remix'' of ancient Greece) but that information technologies have
accelerated these processes and made remixing more salient
\cite{manovich_remix_2005}.  Sinnreich et al. argue that
``configurable culture'' has been significantly transformed by
networked technologies which introduce perfect copying and allow
people not only to be inspired by extant creations but to remix the
original works themselves \cite{sinnreich_ethics_2009}. Legal scholars
have stressed the importance of remixing in cultural creation broadly
and warned that current copyright and intellectual property laws may
hinder creativity and innovation
\cite{lessig_code:_2006,benkler_wealth_2007}.

Several of the most influential scholarly explorations of
remixing as a cultural phenomenon have focused on youth's remixing practices.
For example, work on remixing by Jenkins \cite{jenkins_convergence_2006}
and Ito \cite{ito_personal_2006} has focused on young people's use and
re-use of media. Palfrey and Gasser have suggested that the cultural practices of ``digital native'' 
youth have had a significant transformative effect on our culture \cite{palfrey_born_2008}. 
Throughout his book ``Remix,'' Lessig uses youth's reuse practices to support an
argument against what he considers excessive copyright legal
protection \cite{lessig_remix:_2008}.

\begin{figure}
\begin{center}
\includegraphics[width=1.8in]{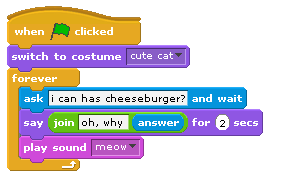}
\caption{An example Scratch program written using programming
``blocks''.}
\label{fig:blocks}
\end{center}
\end{figure}

Yet, despite a wide interest in remixing and authorship, researchers
have only recently engaged in empirical research on the subject
\cite{cheliotis_analysis_2009}. Several recent treatments have
presented studies of video remixing communities
\cite{diakopoulos_evolution_2007,shaw_community_2006}, music remixing
communities \cite{cheliotis_analysis_2009}, collaborative video
game communities \cite{luther_edits_2010} and social network sites \cite{perkel_copy_2006}. 
There is also another quantitative study of our empirical setting \cite{hill_responses_2010} 
focused on characterizing the variety of responses to remixing. 
These studies have tended to be general and
largely descriptive examinations of remixing practice. This work
has pointed to the existence of norms
\cite{diakopoulos_evolution_2007} and the territoriality of digital
creators \cite{thom-santelli_whats_2009} and has considered issues of
motivation \cite{cheliotis_analysis_2009}. However, empirical work has yet to
unpack in detail the key social mechanisms that scholars have
suggested drive behavior, norms, and motivation in remixing
communities.

Perhaps no mechanism has been more frequently cited as critical for
remixing activity than attribution and the related
phenomena of plagiarism, reputation, and status.  For example, recent
survey-based work has suggested that the ``authenticity and legitimacy'' of creative work
``are premised on the explicit acknowledgment of the source materials or
`original creator''' and that such acknowledgment is a key component of how adults assess
the fairness or ethical nature of content reuse
\cite{sinnreich_ethics_2009}.  Attribution, in this sense, can be seen
as an important way that people distinguish remixing from ``theft.''

Judge and law professor Richard Posner stresses the importance of
attribution and explains that this is important even when there is no
monetary benefit to being attributed. For example, he explains that
European copyright law is based on a doctrine of ``moral rights''
that ``entitles a writer or other artist to be credited for his
original work and this `attribution right', as it is called, would
give him a legal claim against a plagiarist.'' Posner also explains
that ``acknowledgment'' of another's contributions to a derivative
negates any charge of plagiarism, although it may not establish
originality \cite{posner_little_2007}. Attribution plays such an
important role in remix culture that Creative Commons made a
requirement for attribution a component of all their licenses after
more than 97\% of licensors opted to require attribution when it was
offered as a choice \cite{brown_announcing_2004}.

Young people's perceptions of attribution and complications around
copying have also been examined. An article by Friedman describes
that adolescents who allowed ``computer pirating'' -- the unauthorized 
copying of computer programs -- did so because technological affordances
made it difficult for adolescents to identify ``harmful
or unjust consequences of computer-mediated actions''
\cite{friedman_social_1997}. In a second study, psychologists Olson
and Shaw have found that by five years old, ``children understand
that others have ideas and dislike the copying of these ideas''
\cite{olson_no_2010}. 

Yet, despite the fact that researchers in human computer interaction
have begun to explore the complexity of attribution and cited its
importance to remixing \cite{luther_edits_2010}, many designers of
online communities pay little attention to issues of attribution
in their designs -- a fact that is reflected in user behavior. For
example, research on the use of photos from
the photo sharing site Flickr \cite{seneviratne_policy-aware_2009}, as well as a number of
other user-generated content communities
\cite{seneviratne_remix_2010}, suggests that most re-users fail to
attribute re-used content in ways that public-use licenses require.
Although theory and survey based work points to a need to design for
attribution in user-generated content communities, we still know very
little about how attribution works or how designers might go about
doing so. Indeed, our study suggests that the most obvious efforts to
design for attribution are likely to be ineffective.

In this paper, we employ a mixed methods approach that combines
qualitative and quantitative analyses to explore users' reactions to
attribution and its absence in a large remixing community. First, we
introduce our empirical setting; using qualitative data from users
forums and comments, we present a rich description of remixing and
evidence to support our core proposition that credit plays a central
role in remixing in our environment. Second, we contextualize and
describe a technological intervention in our setting, responding
directly to several user suggestions, that automated the attribution
of creators of antecedent projects when content was remixed. Third, we
present a tentative quantitative analysis of the effect of this
intervention along with a parallel analysis of the practice of manual
credit-giving. We find that credit-giving, done manually, is
associated with more positive reactions but that automatic attribution
by the system is not associated with a similar effect. Fourth, we
present analysis of a set of in-depth interviews with twelve users
which helps confirm, and add nuance and depth to, our quantitative
findings.

Our results suggest that young users see an important, if currently
under-appreciated and under-theorized, difference between
\emph{credit} and \emph{attribution}. Credit represents more than a
public reference to an ``upstream'' user's contributions. Coming from
another human, credit can involve an explicit acknowledgment, an
expression of gratitude, and an expression of deference, in a way that
simple attribution can not. Our results suggest that identical
attribution information means something very different to users when
it comes from a computer, and when it comes from a human -- and that
users often feel that acknowledgment is worth much less when it comes
from a system. We conclude that designers should create affordances
that make it easier for users to credit each other, rather than to
merely pursue automated means of acknowledgment.

Our study offers two distinct contributions for social scientists and
for technology designers.  The first is an improved understanding of
the way that attribution and credit work in user-generated content
communities. The second is a broader contribution to the literature on
design that suggests an important limitation to technologists' ability to
support community norms and a suggestion for how designers might
create affordances. Functionality that allows users to
express information that a system might otherwise show automatically
may play an important role in successful design for social media
environments.

\section{Scratch: A Community of Young Remixers}

The Scratch online community is a free and publicly available website
where young people share their own video games, animated stories,
interactive art, and simulations
\cite{monroy-hernandez_empowering_2008}.  Participants use the Scratch
programming environment \cite{resnick_scratch:_2009}, a desktop
application, to create these interactive projects by putting together
images, music and sounds with programming command blocks (See Figure
\ref{fig:blocks}).

\begin{figure}
\begin{center}
\includegraphics[width=3in]{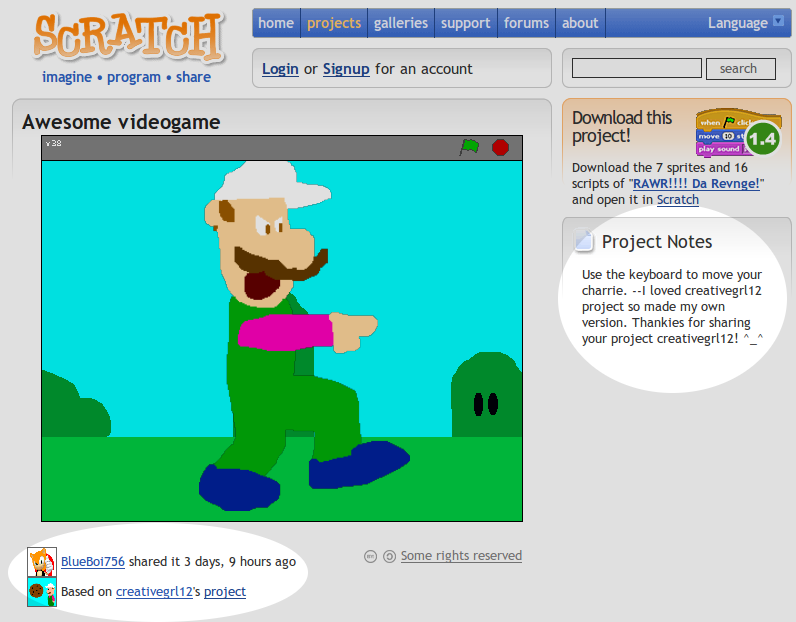}
\caption{Screenshot of a Scratch remix project highlighting automatic
  (the area inside the bottom left circle) and manual (the
  area in the top right circle) attribution.}
\label{fig:projectpage}
\end{center}
\end{figure}

The Scratch website was officially announced in 2007 and, as of
September 2010, has more than 600,000 user accounts who have shared
1.3 million projects. At the time of writing, Scratch users share on
average one new project per minute.  Examples of projects range from
an interactive virtual cake maker, to a simulation of an operating
system, to a Pokemon-inspired video game, to an animation about
climate change, to tutorials on how to draw cartoons.  Like other
user-generated content websites, such as YouTube or Flickr, Scratch
projects are displayed on a webpage (See Figure \ref{fig:projectpage})
where people can interact with them, read metadata and give
feedback. Visitors can use their mouse and/or keyboard to control a
video game or other type of interactive projects or simply observe an
animation play out in a web browser. Metadata displayed next to
projects includes a text-based description of the project, the
creator's name, the number of views, downloads, ``love its,'' remixes,
and galleries (i.e., sets of projects) that the project belongs
to. Users can interact with projects by giving feedback in the form of
tags, comments, or clicks on the ``love it`` button, and can flag a
project as ``inappropriate'' for review by site administrators.

Participants' self-reported ages range primarily from 8 to 17
years-old with 12 being the median.  Thirty-six percent of users
self-report as female.  A large minority of users are from the United
States (41\%) while other countries prominently represented include
the United Kingdom, Thailand, Australia, Canada, Brazil, South Korea,
Taiwan, Colombia and Mexico.  About 28\% of all users -- more than
170,000 -- have uploaded at least one project.

\subsection{Remixing in Scratch}

Scratch users can download any project shared on the website, open it
up in the Scratch authoring environment, learn how it was made, and
``remix'' it. In Scratch, the term ``remixing'' refers to the creation
of any new version of a Scratch program by adding, removing or
changing the programming blocks, images or sounds.  In this section we
use qualitative data from the Scratch website to provide social
context for remixing and to suggest that credit plays an important
role in how users conceive of appropriate remixing practice.

Remixing in Scratch is not only technically possible, it is something
that the administrators of the website encourage and try to foster as
a way for people to learn from others and collaborate.  On every
project page, the Scratch website displays a hyperlink with the text ``Some
rights reserved'' that points to a child-friendly interpretation of
the Creative Commons Attribution-Share Alike license under which all
Scratch projects are licensed.\footnote{A copy of the current version
  of the kid-friendly license is available online at
  http://scratch.mit.edu/pages/license. The version available today
  encourages users to give credit manually in the project notes. A
  strong emphasis on credit-giving was added as a result of the
  findings reported here but was absent during the period of data
  collection for this study.} Even the name Scratch is a reference to
hip hop DJs' practice of mixing records. A large portion of all
projects shared on the Scratch website (28\%) are remixes of other
projects.

That said, remixing is not universally unproblematic in
Scratch. Previous quantitative analysis of the the Scratch community
showed that Scratch participants react both positively and negatively
to the remixing of their projects and found that of those users who
viewed a remix of their project, about one-fifth left positive
comments while the same proportion of users accused the remixer of plagiarism
\cite{hill_responses_2010}.  This ambivalent reaction to remixing is
echoed, and given additional texture, in the comments and complaints
left by users on the Scratch website and sent to Scratch
administrators.

For example, even before the Scratch website was publicly announced,
a number of early adopters became upset when they found remixes of
their projects on the website. Indeed, one of the very first
complaints about Scratch occurred on the discussion forums where a 13
year-old boy asked:

\begin{quote}
  Is it allowed if someone uses your game, changes the theme, then
  calls it `their creation'? Because I created a game called
  ``Paddling 1.5'' and a few days later, a user called ``julie'' redid
  the background, and called it `her creation' and I am really annoyed
  with her for taking credit for MY project!!\footnote{All usernames
    and quotes from the website were changed to disguise the
    identities of participants.}
\end{quote}

A similar complaint was sent to the website administrators by a
14-year old boy:

\begin{quote}
  I think there should be a way to report plagiarized projects I've
  been seeing a lot of people's projects taken and renamed. This
  member, named kings651, has 44 projects, and most of them are made
  by other people. He even has one that I saw my friend make so I know
  he actually made it.
\end{quote}

In other cases, the disagreements over remixing were more public and
involved communication via projects and comments. For example, user
koolkid15 wrote the following message in a comment which was left is
response to a remix that shows a cat frowning:

\begin{quote}
  Hi i'm koolkid15 the original creator of luigi disco jayman41 copied
  me!! and didn't even aknowladge me he didn't change anything !! I
  wrote or drew!!  and jayman...if your reading this think about other
  people!!!!
\end{quote}

Despite the fact that Scratch was conceived, designed, and launched as
a platform for remixing, these users expressed their displeasure at
remixing. That said, none of these users complained directly about the
reuse of their project in general, but in terms of unfair ``taking
credit'', plagiarism, and a lack of acknowledgment. Remixing was seen
as problematic for koolkid15, for example, because of the
non-transformative nature of reuse, the lack of acknowledgment of
antecedent contributors, and the confusion about credit that would
result.

Of course, other, more positive, scenarios around remixing also played
out in Scratch. For example, jellogaliboo created a remix of Catham's
project and wrote the following in the project notes: ``i kinda copied
Catham's "jetpackcat" game. i used the kitty, the blocks (i added and
changed some), and the fuel thingy.'' Catham later posted his approval
of the remix saying, ``I like what you changed about my project!''
Like this example, many of these positive experiences involved
explicit credit-giving by a remixer to the creator of the antecedent
project.

\section{Design Intervention: Automating Attribution}

Several user complaints about remixing and plagiarism also included
suggestions for how Scratch's designers might address them.  For
example, in response to the forum thread mentioned in the previous
section, a 16 year-old proposed two potential design-based solutions:

\begin{quote}
  Make it so you can only download a view of how your
  game/story/animation works.  Or make it so downloadable Scratch
  files have read only protection.  Maybe downloaded Scratch files,
  after being uploaded, are marked with the creators name at the
  bottom, and then any DIFFERENT people who edit it after are put on
  the list.
\end{quote}

Influenced by these comments, Scratch administrators came to believe
that negative responses towards remixing were often due to the fact
that Scratch users did not acknowledge the sources of their
remixes. As a result, these administrators implemented an
architectural design change to the Scratch community along the lines
suggested by the user in the second half of the quotation above.

\begin{figure}
\begin{center}
\includegraphics[width=1.88in]{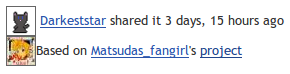}
\caption{Example of the automatic attribution-giving statement
  displayed under remixes in the period after the design
  intervention.}
\label{fig:basedon}
\end{center}
\end{figure}

\begin{figure}
\begin{center}
\includegraphics[width=3.3in]{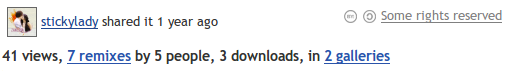}
\caption{Example of the link pointing to remixes of a project
  displayed under antecedent projects in the period after the
  design intervention.}
\label{fig:linktoremixes}
\end{center}
\end{figure}

The design change in question involved the introduction of a new
technological facility that automatically identified and labeled
remixes and inserted hyperlink pointers under each remix to the
remix's antecedent and the antecedent's author (see Figure
\ref{fig:basedon}). Two days after the introduction of this feature,
functionality was added to link to a comprehensive list of derivative
works from the pages of antecedent projects (see Figure
\ref{fig:linktoremixes}).

The new feature was announced in the discussion forums by an
administrator of the website and user responses were
positive. User terminator99 suggested that the change was,
``Awesome.'' Another user, marsUp, posted a comment saying, ``That's a
very useful feature! I like that we can do ping-pong like modding in
Scratch.'' Users who did not visit the discussion forums also
responded well to the the new feature. For example, user greekPlus
posted a comment on a remix he created saying, ``i remixed it for you
but i do not know how to ad credit to you for thinking of it in the
first place.'' A few minutes later he realized that the remix
automatically displayed the attribution and posted the a comment
saying, ``never mind it did it for me. cool!''

\section{Study 1: Human and Machine Attribution}

Although initial user feedback to the automatic attribution feature
was positive, users continued to complain about remixing.  In Study
1a, we present a quantitative analysis to more fully evaluate the
effect of the technological design change described in the previous
section.  In Study 1b, we offer a parallel analysis of the
relationship between manual crediting-giving by users and users'
reactions to being remixed.

Both studies build on a dataset used in previous work by Hill,
Monroy-Hernández, and Olson \cite{hill_responses_2010}. This dataset
includes remix-pairs determined by an algorithm using detailed project
metadata tracked by the Scratch online community.  The dataset is
limited in that it does not include projects whose concepts were
copied by a user who had seen another’s work but who did not actually
copy code, graphics or sound. Similarly, the dataset contains no
measure of the ``originality'' of projects or an indicator based on
ideas that were taken from a source outside Scratch (e.g., a user may
have created a Pacman clone which would not be considered a remix in
our analysis).

The data presented here includes each coded reactions of the author of
antecedent projects (i.e., originators) on remixes of their projects
shared by other users in the site during a twelve week period after
Scratch's launch from May 15 through October 28, 2007. Although 2,543
remixes were shared in this period, we limit our analysis to the 932
projects (37\% of the total) that had been viewed at the time of data
collection by the project originator -- a necessary prerequisite to
any response. Of these 932 remixes that were viewed by a project
originator, 388 originators (42\%) left comments on the remixes in
question. The remaining were coded as ``silence.''  Comments left by
originators were coded by two coders, blind to the hypotheses of the
study and who were found to be reliable \cite{hill_responses_2010}, as
being positive, neutral, or negative. They were also coded as 
containing accusations of plagiarism (projects in which the the
originator directly accused the remixer of copying, e.g., ``Hello mr
plagiarist'', ``Copy-cat!'')  or hinting plagiarism (projects in which
the originator implied that the remixer had copied but did not state
this explicitly, e.g., ``I mostly pretty much made this whole entire
game'').

Unless it also contained an explicitly negative reaction, an
accusation of plagiarism was not coded as ``negative.''  However,
because plagiarism tends to be viewed as negative within
Scratch (as suggested by the quotations in the previous section) and
more broadly in society \cite{posner_little_2007}, we re-coded
accusations of plagiarism (both direct and hinting) as ``negative''
except, as was the case in several comments coded as ``hinting
plagiarism,'' when these accusations were in comments that were also
coded as positive. Previous published work using this dataset, and
subsequent robustness checks, show that our results are substantively
unchanged if we exclude these explicit charges of plagiarism from the
``negative'' category or exclude only the weaker ``hinting
plagiarism'' accusations.

\subsection{Study 1a: Automatic Attribution}

\begin{figure}
\begin{center}
\includegraphics[width=3.3in]{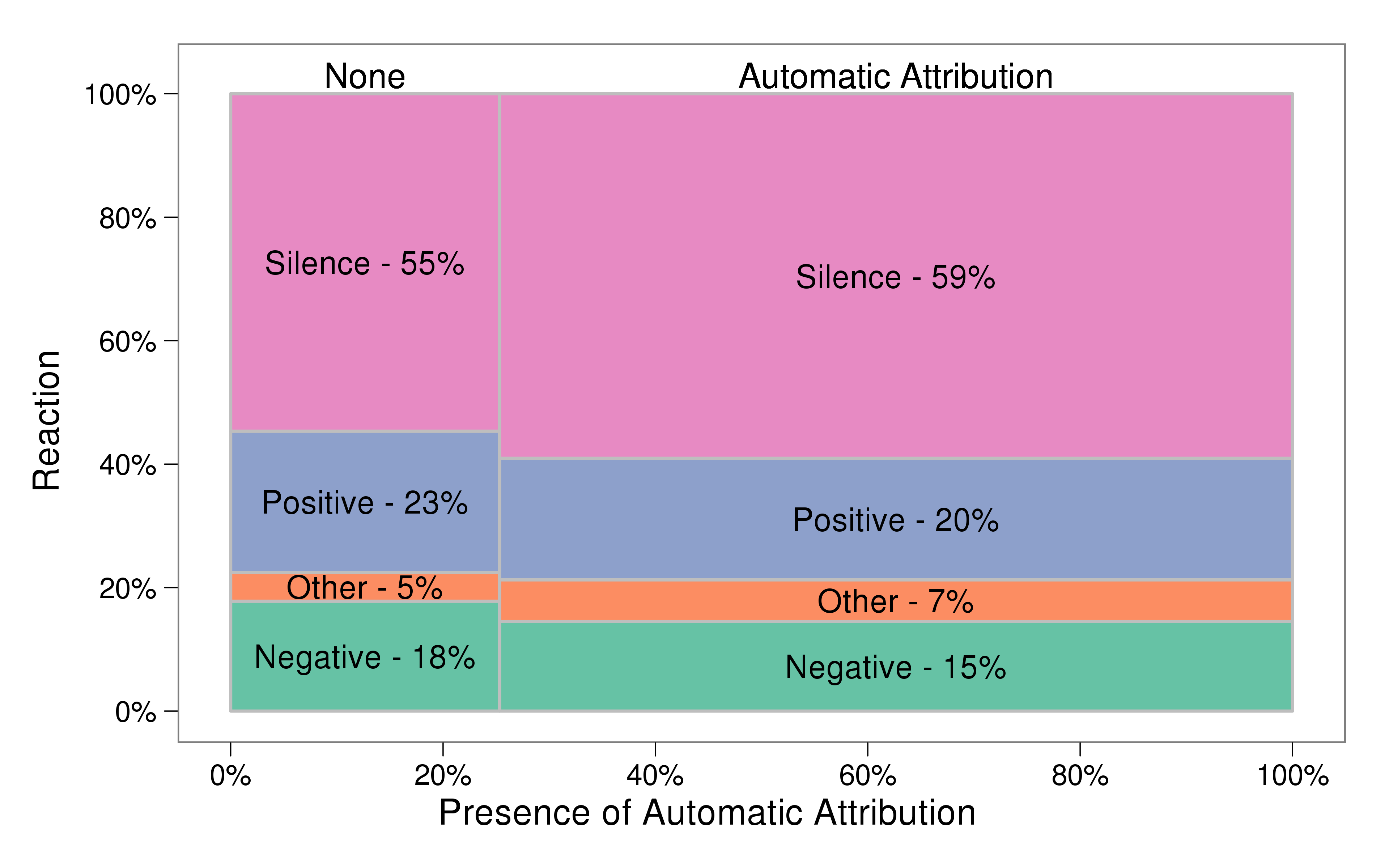}
\caption{Mosaic plot showing the distribution of reactions of original
  creators who had viewed remixes of their project during the six week
  periods before and after the introduction of automatic
  attribution. The proportion of response types is shown along the
  y-axis. The proportion of projects viewed with, and without,
  automatic attribution is shown along the x-axis. ($n=932$)}
\label{fig:mosauto}
\end{center}
\end{figure}

To test the effectiveness of automatic attribution, we consider the
effect of the design intervention described in the previous section.
The design change took place six weeks after the public launch of the
Scratch community and at the precise midpoint in our data collection
window. The intervention affected all projects hosted on the Scratch
online community including projects shared before the automatic
attribution functionality was activated. As a result, we classify
originators' reactions as occurring outside a technological regime of
automatic attribution when a project was both uploaded and viewed by a
project's originator before automatic attribution functionality was
activated.

A comparison of the distribution of coded comments between positive,
neutral, negative, and silent in the periods before and after the
intervention suggests that the introduction of automatic attribution
had little effect on the distribution of reaction types (See Figure
\ref{fig:mosauto}). Although the period after the intervention saw a
larger proportion of users remaining silent and a smaller proportion
of both positive and negative comments, $\chi^2$ tests suggest that
there is no statistically significant difference in originator
reactions between remixes viewed before or after the introduction of
automatic attribution ($\chi^2 = 3.94; df=3; p=0.27$). As a result, we
cannot conclude that the there is any relationship between the
presence, or absence, of an automatic attribution system in Scratch
and the distribution of different types of reactions.

These results suggest that automatic attribution systems may have
limited effectiveness in communities like Scratch. Of course, our
analysis is not without important limitations. For example, the
existence of an automatic attribution regime may also affect the
behavior of users preparing remixes. A remixer might avoid making
perfect copies of projects if they know that their copies will be
attributed and are more likely to be discovered.

\subsection{Study 1b: Manual Crediting}

While the introduction of an automatic attribution feature to Scratch
appears to have had a limited effect on originators responses to
remixes of their projects, the presence or absence of credit was a
recurring theme in discussions on Scratch online forums -- as shown in
the quotes in the previous section -- and in many of the coded
reactions from the periods both before and after the introduction of
automatic attribution. Indeed, in project descriptions or notes from
the periods both before and after the change, remixers frequently
``manually'' gave credit to the originators of their
work. Even after remixes were automatically attributed to originators,
remixers who did not also give credit manually -- essentially
producing information redundant to what was already being displayed by
the system -- were criticized.

For example, after the introduction of automatic attribution
functionality, a user left the following comment on a remix of their
project:

\begin{quote}
  Bryan, you need to give me Pumaboy credit for this wonderful game
  that I mostly pretty much kinda totally made this whole entire game
  ... and that you need to give me some credit for it
\end{quote}

For this user, automatic attribution by the system did not represent a
sufficient or valid form of credit-giving. In the following study, we
test for this effect of ``manual'' credit-giving by remixers on coded
response types using a method that parallels the analysis in Study 1a
and that uses the same dataset.

Manual crediting can happen in multiple ways.  Exploratory coding of
133 randomly selected projects showed that 35 (36\%) of each remix pair gave
credit.  Of these 35 projects, 34 gave credit in the project
description field while 1 project only gave credit in a ``credits''
screen inside the game. As a result, the authors of this study split
the sample of projects used in the Study 1a and coded each of of the
user-created descriptions for the presence or absence of explicit or
manual credit-giving.

To first establish that we are examining distinct behaviors, we
attempted to establish that automatic and manual attribution do not
act as substitutes for each other. As suggested by our qualitative
findings and our results in Study 1a, we found little difference in
the rate of explicit credit giving between projects created in the
presence or absence of automatic attribution. Overall, 276 (about
30\%) of the 932 projects in our sample offered explicit credit in the
description field of the project. Manual crediting-giving was a
widespread practice both before automatic attribution, when 31\% of
projects in our sample offered explicit credit, and after, when 27\%
did so.  The difference between these two periods was not
statistically significant ($\chi^2=1.41; df=1; p=0.24$). Previous work
studying \emph{Jumpcut}, a video remixing website, supports the idea
that automatic and manual credit giving are not interchangable
phenomena.  One Jumpcut user with permission to creative derivative
works commented that they, ``still feel a moral obligation to people
as creators who have a moral right to be attributed (and notified)
despite the physical design which accomplishes this automatically''
\cite{diakopoulos_evolution_2007}.

\begin{figure}
\begin{center}
\includegraphics[width=3.3in]{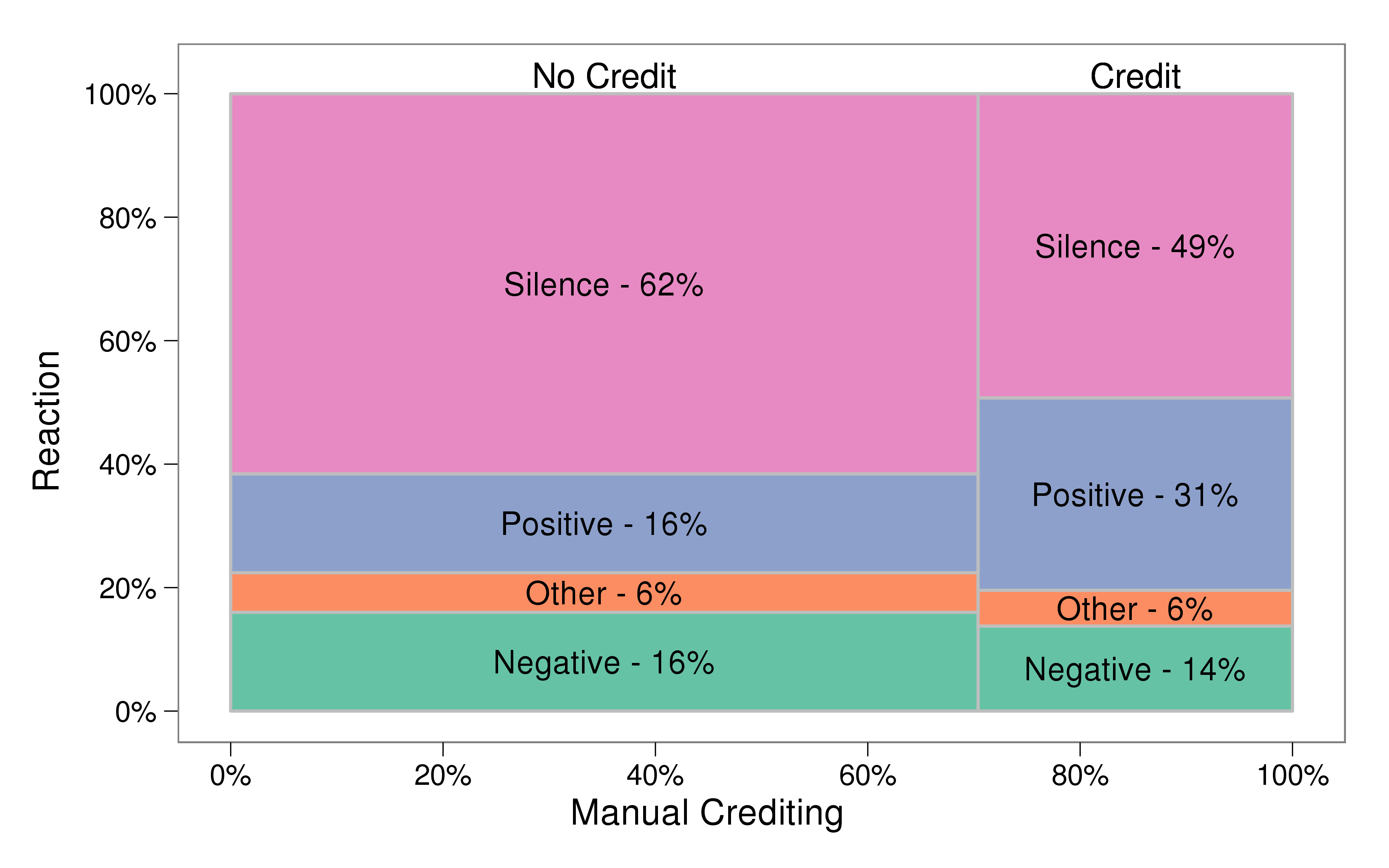}
\caption{Mosaic plot showing the distribution of reactions of original
  creators who had viewed remixes of their project and indicating
  whether they left manual credit. The proportion of response types is
  shown along the y-axis. The proportion of projects including and
  omitting manual credit is shown along the x-axis. ($n=932$)}
\label{fig:moscredit}
\end{center}
\end{figure}

We measured effectiveness of manual credit giving using a parallel
analysis to Study 1a. As in Study 1a, we compared the distribution of
originator reactions in the presence, and absence, of manual
credit-giving by remixers. We found that negative reactions are less
common in the presence of manual credit but that this difference is
very small (from 16\% without manual credit to 14\% with it). However,
we see that the proportion of users who react positively almost
doubles in the presence of credit-giving (from 16\% with no crediting
to 31\% in its presence). A graph of these results are shown
in Figure \ref{fig:moscredit}. Tests show that we can confidently
reject the null hypothesis that these differences in the distribution
of reactions are due to random variation ($\chi^2 = 27.60; df=3;
p<0.001$).

Also important to note is a difference in the number of users who are
silent after viewing a project (62\% in the absence of manual credit
versus 49\% in its presence). This larger proportion of commenting in
general may have an important substantive effect on the discourse and
behavior on the site because silent originators may, for obvious
reasons, have a more limited effect on attitudes toward remixing and
user experience than vocal users do. As a robustness check, we
considered the reaction of only originators who left comments
($n=388$) and found that even with a smaller sample, our result were
stronger. In the restricted sample, 41\% reacted negatively when they
were not given credit.  However, only 27\% did so when they were
credited. Similarly, 42\% of users who left comments on projects that
did not give credit manually left positive messages. Nearly two thirds
of comments (61\%) were positive when credit was given.  These
differences, in the reduced sample that includes only explicit
reactions, were also statistically significantly different ($\chi^2 =
14.09; df=2; p<0.001$). We include the large number of silent
participants because we believe that non-response is an important type
of reaction with real effects on the community. Understanding the
reasons behind non-response and the effect of silence in response to
different types of credit giving remains an opportunity for further
research.

Although not presented here due to limited space, we followed the
general model of previous work using this dataset
\cite{hill_responses_2010} and tested logistic regression models on
dichotomous variables indicating the presence of negative and positive
reactions and found that basic relationships described above were
robust to the introduction of a control for the the intervention, to
an interaction between these two variables, and to controls for the
gender and age of originators and to the antecedent project's
complexity. Both before or after the intervention, manual crediting
resulted in more positive comments by the originators of remixed
projects. Of course, the results presented here are uncontrolled,
bivariate, relationships and we caution that these results, while
provocative, should still be viewed as largely tentative. As we show
in the subsequent qualitative analysis, attribution and credit-giving
are complex social processes. We do not claim that the preceding
analyses capture it fully.

\section{Study 2: Interviews with participants}

In order to explore the reasoning behind young people's remixing
behavior and attitudes toward attribution as we observed it in Study
1, we engaged in a second qualitative study and directly asked kids
what role attribution and credit plays in their moral evaluations of
remixing.

\begin{table*} \begin{tabular}{lccp{4.5in}} \hline
\textbf{Name (pseudonym)} & \textbf{Age} & \textbf{Gender} & \textbf{Relationship to Scratch}\\
\hline
Nicole & 10 & F & She has created with hundreds of Scratch projects, mainly animations and art ones. \\
Kyle & 14 & M & Casual user of Scratch, interested in math/science simulations and video games. \\
Amy & 15 & F & Avid photographer, has never used Scratch. \\
Charles & 9 & M & Active member of a subgroup of Scratch interested in simulation of operating systems.  \\
Ryan & 12 & F & Long-time member of the Scratch community. Creates complex video games. \\
Jon & 9 & M & Casual user of Scratch, collaborates with Scratch friends in person. \\
Jake & 11 & M & Casual user, likes making video games. \\
Cody & 16 & M & Creates hip hop accessories, not active in Scratch. \\
Paul & 9 & M & Creates Scratch projects with a focus on engineering and video games. \\
Jimena & 17 & F & Highly technical teen with programming experience but no experience with Scratch. \\
Madeline & 14 & F & Very popular animator  in the Scratch community. \\
Susie & 10 & F & Has created hundreds of projects including games, animations and art, but preferring art. \\
\hline
\end{tabular} \caption{Table listing details of interviewees used in Study 2. ($n=12$)} \label{tab:ints} \end{table*}

\subsection{Methodology}

We conducted twelve one-hour semi-structured interviews with kids aged
8 to 17 years old. All of the interviewees had 
experience using
computers and had access to the Internet at home. All the interviewees
live in the United States except for one who lives in New Zealand. The
participants were recruited via the Scratch website and during
meet-ups with educators, teachers and young Scratch users. Eight of
the interviews were conducted in person, in the Boston area, and the
rest over the phone or voice over IP.  The interviews were
audio-recorded and transcribed before fully analyzing them. Nine of
the interviewees were members of the Scratch community.  The remaining
three did not use Scratch but were included as a way to check if
people who do not use Scratch have similar views about remixing,
attribution, and credit. We found no substantive difference between
the Scratch users and non-users in their answers to questions related
to the hypothetical automatic and manual mechanism for attribution.

Before each interview, subjects completed a survey that elicited
demographic information and posed questions about their familiarity
with other technologies and which was primarily designed to get a
sense of the interviewees' social and technical background. Interviews
were structured around a protocol that included a set of nine
fictional remixing cases intended to elicit conversations about
remixing.\footnote{Our interview protocol including example cases is
  available at
  http://www.media.mit.edu/~andresmh/chi2011/interview.html.} The
cases were inspired by Sinnreich et al.'s theoretical work and from
three years of experience moderating the Scratch community. They were
designed to present cases where remixing could be controversial but
where there is no clear ``correct'' answer. The goal of the cases was
to offer a concrete, and common, set of dilemmas to stimulate broad
conversations about attitudes toward remixing.

The cases were presented in the form of printed screenshots of
different project pages from the Scratch website (anonymized to avoid
referring to real cases that users might have seen).
The print outs were shown to the interviewees (or
discussed over the phone) while explaining each case.  All the cases
included a remix and its corresponding antecedent project. The cases varied 
in the presence of automatic attribution, manual credit, and
the degree of similarity between the remix and its antecedent. For example, the
first three cases were:

\begin{enumerate}

\item A remix and its antecedent are identical. The project notes only
  describe how to play the video game. The remix shows the automatic
  attribution but no manual credit on the notes.

\item A remix and its antecedent are different (as seen visually and
  in project metadata) but one can clearly see the influence of its
  antecedent project. The project notes of the remix show manual
  credit but no automatic attribution. The interviewee was told to
  imagine the site had a glitch that prevented it from connecting it
  to its antecedent.

\item The same set of remix and antecedent projects as in (2) but this
  time automatic attribution is displayed but manual credit is not.

\end{enumerate}

Each of the interview logs was coded using
inductive codes and grounded theory \cite{charmaz_constructing_2006}.
The coded responses were analyzed based on categories related to how
interviewees answered specific questions about the distinction between
automatic attribution and manual credit.

\subsection{Results}

Confirming the results of Study 1, for users of Scratch, automatic
attribution was generally seen as insincere and insufficient.
Throughout the interviews, we found that for most of the kids, getting
explicit credit from another person was preferred over attribution
given automatically by the system. When asked why, kids often
responded that knowing that another person had cared enough to give
credit was valued more than what the computer system would do on its
own. The fact that it takes some work, albeit minimal, to write an
acknowledgment statement, sends a signal of empathy, authenticity and
good intentions \cite{donath_signals_2008}. Amy articulated this when
explaining why she preferred getting credit from another person:

\begin{quote}
  I would like it even more if the person did it [gave credit] on their own accord,
  because it would mean that [...] they weren't
  trying to copy it, pirate it.
\end{quote}

Similarly, Jon explained, ``No [the ``Based on'' is not enough],
because he [the remixer] didn't put that, it always says that.'' For
Jon, automatic attribution is not authentic because it is always there
and, as a result, it is clear that is not coming from the person doing
the remix.

Most of the interviewees seemed to have a clear notion of what they
think a moral remix should be. For some, it is all about making
something different. Jake for example, defines a ``good'' remix as,
``if it has a bunch of differences then it’s a good remix. If it has
like two, then it’s bad.''  In addition to the differences between the
remix and its antecedent project, for some, manual credit is part of
what makes it moral. Charles said, ``[remixing] is taking somebody
else's project and then changing a lot of it and sharing it and giving
credit.''  Continuing, Charles explained:

\begin{quote}
  If Green had actually said in the project notes, ``This is a remix
  of Red's project, full credit goes to him,'' then I would consider
  it a remix.  But this [pointing at a remix without manual credit] is
  definitely a copy.
\end{quote}

Likewise, Ryan mentions that a fictional remix was, ``perfectly fine
because they gave credit in the project notes.''

Interviewees suggested that manual credit also allows users to be more
expressive. For example, Susie explained that expressiveness is the
reason that she prefers manual credit through the project notes
saying, ``I think the manual one is better because you can say `thank
you' and things like that.  The automatic one just says `it's based
on.''' Susie also notes that for her, the project notes are a space
where a creator can express her wishes in regards to her intellectual
property, independent, and even in contradiction to, the license of
the projects:

\begin{quote}
  If I do a project that has music that I really like, I often
  download the project, take the music.  Unless it says in the project
  notes, ``Do not take the music.''
\end{quote}

For Susie and other users of Scratch, the project notes are a space
for more than just instructions on how to interact with one's project;
they are an expressive space where one can communicate with an
audience without having to encumber the creative piece of work with
it.

Others point at the fact that people do not pay as much attention to
automatic attribution statement as much they do to the manual credit
left in project descriptions. Jake, for example, explains that, while
he agrees there is some usefulness to having both, project notes still
are more important, ``because, you know, sometimes people just like
skim through a project and you don’t see it ‘til the end.''  Jake
continued to say that creators that do not have both should get a
``warning.''

Even though interviewees value manual credit, they still see the
usefulness of the automatic mechanism as some sort of
community-building prosthetic device -- an explanation for the
positive reactions to the feature's initial introduction. For example,
Nicole argues that while manual credit on the notes has more value for
her, the automatic attribution is useful as a backup and because it
provides a link:

\begin{quote}
  Well, I think that they should probably write in the notes that --
  then it should also say ``Based on blank's project,'' just in case
  they forget, and also because it gives a link to the original
  project and it gives a link to the user so you don't have to search
  for it.
\end{quote}

A similar explanation was articulated on a comment exchange on one the
website's galleries. A teenage girl that actively participates in
Scratch explained the pragmatic value of automatic attribution saying,
``the `based on' thingy, it gives a link, and we all luv links, less
typing,'' before reiterating that manual credit is more valuable:

\begin{quote}
  at the beginning i thought that you don't have to give credit when
  the ``based on'' thingy is in there, but i realized a lot of people
  don't look at that, and i noticed people confused the remix with the
  original.
\end{quote}


Creating a Scratch project is a complicated task. A project's sources
can be diverse and the creator can easily forget to acknowledge some,
as Paul explains, when asked to choose between a system of manual
credit or automatic attribution:

\begin{quote}
  The thing is, it would be a lot better if they had both. Because,
  sometimes people probably just forget to do that. And then people
  would not know.
\end{quote}

There are also situations where interviewees recognize what Posner
calls the ``awkwardness of acknowledgment,'' that is, situations where
credit is not really needed and it can be an unnecessary burden or go
against the aesthetics of the work \cite{posner_little_2007}. For
example, Paul mentioned that sometimes, there are some projects in
Scratch that are remixed so much -- like the sample projects that come
with Scratch or some ``remix chains''\footnote{Remix chains typically
  start with someone sharing a project inviting others to remix 
  (i.e. ``add your animated avatar to the park.'')} -- where credit is not
necessary:

\begin{quote}
  There's this one called ``perfect platformer base'' which a lot of
  people remix. So I don't think that needs any credit. It's not
  actually a real game. It's all the levels and stuff are just
  demonstrations.
\end{quote}

Since manual crediting has a higher emotional value, some kids
mentioned that conflicts over remixing could be addressed by the
administrators of the site by editing the project of the remix in
question, as a way to enforce credit without transforming it into
attribution. Doing so would make it appear that a remixer had credited
an antecedent when they had not. Susie offers a suggestion along these
lines when asked about how the administrators of the website should
deal with a case of a complaint over a remix that is a parody of
someone else's project. Susie suggested that, ``I might remove the
project but I might not, you know, maybe I would edit the notes to to
give credit.''  Similarly, Charles described his approach for solving
conflicts if he was the administrator of the website suggesting that,
``I probably just would stay out of the argument.  I probably wouldn't
remove it [the remix], I'd just add something in the project notes
[like] `based on Gray's project.'''

This phenomena of giving less value to technologically simplified
social signals is experienced in other social platforms. For example,
Amy expressed how on the social network site Facebook, she loves to
get comments on her photographs but dislikes those who do not leave
comments or opt instead to press the ``I like it'' button:

\begin{quote}
  I love when people comment on my pictures.  Everybody sees them,
  because they tell me they have.  I'm like, ``Oh really?  That's
  great.  Why didn't you comment?'' I don't like it when people just
  ``like it'', because you know they have something to say about it;
  they just don't.  It's like, if they like it, then [they should]
  take the time to say something.
\end{quote}

Although not designed to be a random sample, these interviews support
the proposition that both Scratch participants and other young people
share a set of norms about characteristics that determine what a
``good'' or moral remix is. Among these norms, acknowledging one's
sources seems to play a central role. However, participants also seem
to share the opinion that this norm is not satisfied through an
automated process.  They clearly understand the pragmatic value of
automating acknowledgment-giving, but
they do not see it as a substitute for adherence to the social norm of
credit-giving. They also see it as void of emotion and expressiveness.
For Scratch users, normative constraints are separate from
architectural constraints and one cannot replace the other.
These findings support and enrich the results from our first
study and help us understand better how Scratch participants, and
perhaps kids in general, experience authorship norms and automation in
online spaces.

\section{Conclusions}

Our results from Study 1a called into the question the effectiveness
of automatic attribution functionality in encouraging more positive
user reactions in Scratch. We build on these results in Study 1b to
suggest that manual crediting may do the work that Scratch's designers
had hoped automatic attribution would. Results from the analysis of
user interviews presented in Study 2 help to answer the question of
``why?'' and suggest that users find manual credit to be more
authentic and more meaningful to users because it takes more time and
effort. Usually, UI  improvements are designed to help reduce the time
and effort involved in using a system. But in trying to help users by attributing automatically,
Scratch's designers misunderstood the way that attribution as a social
mechanism worked for Scratch's users. Our fundamental insight is that
while both attribution and credit may be important, they are distinct
concepts and that credit is, socially, worth more. A system can
\emph{attribute} the work of a user but \emph{credit}, which is seen
as much more important by users and which has a greater effect on user
behavior, cannot be done automatically. Computers can
attribute. Crediting, however, takes a human.

As we suggested in our introduction, this fundamental result leads to
two distinct contributions. First, and more specifically, our analysis
offers an improved understanding of the way that attribution and
credit works in user-generated content communities over what has been
available in previous work. Our two studies suggest that scholars are
correct to argue that credit plays an important role in social media
communities and offer empirical confirmation for the important role
that authenticity plays in how users conceptualize credit. In our
in-depth interviews, we explain some of the reasons why this may be
the case. Second, through our evaluation of an unsuccessful
technological design, our work offers a broader, if more preliminary,
contribution in suggesting an important limit of designers' ability to
support community norms in social media systems. As the literature on
design and social media grows, the importance of good support for
communities with healthy norms promoting positive interactions is
likely to increase. In attempting to design for these norms, we
suspect that researchers will increasingly encounter similar
challenges.

We argue that designers should approach interventions iteratively.
This design approach can be understood through the
theoretical lens of the social construction of technology
\cite{pinch_social_1984}: designers can't control technological outcomes which
must be built through a close relationship between designers and users.
Designers must move away from seeing their profession as
providing solutions. They must channel users, work closely with them,
and iterate together, to negotiate and achieve a set of shared goals.

The prevalence of user-generated content sites stresses the importance
of how online social spaces should deal with issues of attribution and
our results are likely to be immediately relevant to 
designers. For example, the Semantic Clipboard is a tool built as a
system of automatic attribution for content reuse
\cite{seneviratne_policy-aware_2009}. Developed by researchers who
found a high degree of Creative Commons license violations around the
re-use of Flickr images, the tool is a Firefox plugin that provides,
``license awareness of Web media,'' and enables people to
automatically, ``copy [media] along with the appropriate license
metadata.'' Our results suggest one way that this approach
may fall short.

However, automatic attribution is not the only way that technologists
can design to acknowledge others' contributions. Indeed, our results
suggest that there may be gains from design changes which encourage
credit-giving without simply automating attribution. For example,
Scratch's designers might present users with a metadata field that
prompts users to credit others and suggests antecedent authors whose
work the system has determined may have played a role. This affordance
might remind users to credit others, and might increase the amount of
crediting, while maintaining a human role in the process and the extra
effort that, our research has suggested, imbues manual credit giving
with its value. We suggest that in other social media communities,
similar affordances that help prompt or remind users to do things that
a system might do automatically represent a class of increasingly
important design patterns and a template for successful design
interventions in support of community norms.


\section{Acknowledgments}

This research was supported by Microsoft Research.
Scratch is a project of the Lifelong Kindergarten Group at the MIT
Media Lab with financial support from the National Science
Foundation award ITR-0325828, Microsoft Corp., Intel Foundation,
Google, the MacArthur Foundation and the MIT Media Lab research
consortia.

\bibliographystyle{acmnocaps}
\bibliography{chi2011}

\end{document}